\documentclass[traditabstract,letter,proofs]{aa} 
\usepackage{graphicx}
\usepackage{latexsym}
\usepackage{color}
\usepackage{natbib}
\bibpunct{(}{)}{;}{a}{}{,} 
\usepackage{longtable,lscape}

\newcommand{\teff}{$T_{\rm eff}$}
\newcommand{\logg}{$\log g$}
\newcommand{\vsini}{$v\sin i$}

\newcommand{\feh}{[Fe/H]}

\newcommand{\ROTFIT}{{\sf ROTFIT}}

\newcommand{\kms}{km\,s$^{-1}$}

\definecolor{blu}{rgb}{0,0,1}
\definecolor{mag}{rgb}{1,0,1}

\begin{document}

\title{ISO-ChaI\,52: a weakly-accreting young stellar object with a dipper light curve.\thanks{Based on observations collected 
at the ESO REM telescope (La Silla, Chile) and at the ESO VLT (ID 084.C-1095). }}

\author{A. Frasca\inst{1} 
\and    C. F. Manara\inst{2}
\and    J. M. Alcal\'a\inst{3}
\and    K. Biazzo\inst{4}
\and    L. Venuti\inst{5,6}
\and    E. Covino\inst{3}
\and    G. Rosotti\inst{7}
\and    B. Stelzer\inst{8,6}
\and    D. Fedele\inst{9}
       }

\institute{INAF -- Osservatorio Astrofisico di Catania, via S. Sofia 78, 95123 Catania, Italy\\ \email{antonio.frasca@inaf.it}
\and
European Southern Observatory, Karl-Schwarzschild-Strasse 2, 85748 Garching bei M\"unchen, Germany 
\and
INAF -- Osservatorio Astronomico di Capodimonte, via Moiariello, 16, 80131 Napoli, Italy
\and
INAF -- Osservatorio Astronomico di Roma, Via Frascati 33, 00078 Monte Porzio Catone, Italy
\and
NASA Ames Research Center, Moffett Blvd, Mountain View, CA 94035, USA
\and
INAF -- Osservatorio Astronomico di Palermo,  Piazza del Parlamento 1, 90134 Palermo, Italy
\and
Leiden Observatory, Leiden University, P.O. Box 9513, NL-2300 RA Leiden, the Netherlands
\and
Institut f\"ur Astronomie und Astrophysik, Eberhard-Karls Universit\"at T\"ubingen, Sand 1, 72076 T\"ubingen, Germany
\and
INAF -- Osservatorio Astrofisico di Arcetri, Largo E. Fermi 5, 50125, Firenze, Italy
}

\date{Received 14 April 2020 / Accepted 12 June 2020}

\abstract{We report on the discovery of periodic dips in the multiband lightcurve of ISO-ChaI\,52, a young stellar object in the Chamaeleon\,I dark cloud.
This is one among the peculiar objects that display very low or negligible accretion both in their UV continuum and spectral lines, although
they present a remarkable infrared excess emission characteristic of optically-thick circumstellar disks. 
We have analyzed a VLT/X-Shooter spectrum with the tool ROTFIT to determine the stellar parameters. The latter,
along with photometry from our campaign with the REM telescope and from the literature, have allowed us to model the spectral energy distribution 
and to estimate the size and temperature of the inner and outer disk. From the rotational period of the star/disk system of 3.45 days we estimate  
a disk inclination of 36$\degr$.
The depth of the dips in different bands has been used to gain information about the occulting material. A single extinction law is not able to fit the observed
behavior, while a two-component model of a disk warp composed of a dense region with a gray extinction and an upper layer with an ISM-type extinction 
provides a better fit of the data.  } 

   \titlerunning{ISO-ChaI\,52: a weakly-accreting YSO with a warped disk}
      \authorrunning{A. Frasca et al.}

\maketitle

\section{Introduction}
\label{Sec:intro}

 A key issue in the study of planet formation is to explain how optically thick accretion
disks surrounding the youngest solar-mass stars and giving rise to a remarkable infrared (IR) excess
(class\,II IR sources) evolve into optically thin debris disks, passing through a phase where only a mild (or null) 
IR excess is visible (class\,III sources).
Normally, thick disks are observed around the classical T Tauri (CTT) stars, which display strong 
emission lines produced by mass accretion, while the  weak-line T Tauri 
(WTT) stars, with negligible signatures of accretion, very often show up as class\,III sources  \citep[][and references therein]{Hartmann2005, Lada2006}.
 Generally, these disks persist for a few million years, 
 during which part of the material is accreted onto the star, part is lost via outflows and
 photoevaporation \citep[e.g.,][]{Ercolano2017,Nisini2018}, 		
and part condenses into centimeter-sized and larger bodies or planetesimals \citep[e.g.,][]{Testi2014}.
 A possible intermediate stage of T Tauri disk evolution is observationally identified with the so-called
 {\it transitional disks} (TDs), which are characterized by inner holes and gaps in their dust distribution 
\citep[e.g.,][and references therein]{Espaillat2014}.

A further category of objects is recently emerging with apparently very little or no evidence for accretion in optical ($\lambda>$3400\,\AA) spectra, 
yet with the near-infrared (NIR) emission characteristic of optically thick dust in the inner (few AU) regions of the disk, so that their spectral energy distribution 
looks like that of class II sources \citep[e.g.,][]{Wahhaj2010,Alcala2019, Thanathibodee2019}. The existence of such objects
might be explained by slightly different timescales for the decline of disk and accretion processes in 
young stars \citep[e.g.,][]{Fedele2010}.  Another possibility is that the accretion is highly variable and occurs mainly during bursts \citep[e.g.,][]{Cody2017}.
Weak accretion, in general, is not easily detectable in the region of the Balmer jump, hence other diagnostics like modeling of the H$\alpha$ line profile 
\citep[e.g.,][]{Espaillat2008, Thanathibodee2019} and/or measures of excess emission in near-UV/far-UV  spectra \citep[e.g.,][and references therein]{Alcala2019} 
 are necessary in these cases of very low accretion rates.

Recently, in our X-Shooter surveys of the Chamaeleon~I \citep[Cha~I,][]{Manara2016, Manara2017} and Lupus star forming regions 
\citep{Alcala2014, Alcala2017}, we have detected several such weak accretors. Thereby we have confirmed that some of these disk-bearing objects lack the 
X-Shooter continuum UV excess, typical of accreting objects. The recent ALMA surveys of \citet{Pascucci2016} in Cha~I and of \citet{Ansdell2016b} in 
Lupus show that the disks around some of these objects are bright in the sub-mm and still host a large amount of dust.

YSOs display luminosity variations on different timescales due to geometric and intrinsic effects. 
The variability of non-accreting objects is mainly related to stellar magnetic activity 
(cold photospheric spots, flares, etc.). For accreting objects, a variety of processes, including hot spots, variable 
circumstellar extinction and/or burst of accretion, can be at the origin of such variations.  
A way to efficiently characterize such processes is  by  photometric multiband imaging techniques. 
Long-term ground-based observations \citep[e.g.,][]{Herbst2002, Frasca2009} and, more recently, 
high-precision, high-cadence space photometry  \citep[e.g.,][]{Venuti2017, Stauffer2017},  
enabled the exploration of different scenarios for the physical process involved in the variability 
of YSOs.	An intriguing observed  behavior is the presence of recurrent 
luminosity dips that are likely due to the periodic occultation of the central star by the magnetically-warped 
inner disk edge \citep[e.g.,][]{Bouvier1999,Bouvier2003, McGinnis2015}.	

In this letter we report on the discovery of quasi-periodic luminosity dips in the weakly accreting object \object{ISO-ChaI\,52}
(= BYB\,18 = 2MASS\,J11044258-7741571)
that were observed simultaneously from the optical to the NIR with the Rapid Eye Mount (REM) telescope 
at the La Silla observatory.
We have also analyzed an intermediate-resolution spectrum of this source ($R\simeq 18\,000$) taken on 18 Dec 2010 
with X-Shooter at the ESO VLT (program ID 084.C-1095) with the aim of deriving stellar parameters in support of this study.

\section{Observations}	
\label{Sec:Observations}

The photometric observations were performed with the 60-cm robotic REM telescope  located at the ESO-La Silla Observatory (Chile), 
on 80 nights from  3 April to 1 October 2019. 
By means of a dichroic, REM feeds simultaneously two cameras at the two Nasmyth focal stations, one for the NIR (REMIR) and one for 
the optical (ROSS2). The cameras have nearly the same field of view of about $10\arcmin\times 10\arcmin$ and use wide-band filters 
($J$, $H$, and $K'$ for REMIR and Sloan/SDSS $g'$, $r'$, $i'$, and $z'$ for ROSS2). 
Due to a technical failure in the REMIR camera, we have NIR data only for the first 20 nights of the campaign.
In total, we collected 260, 277, 280, 285, 103, 102, and 108 usable images in $g'$, $r'$, $i'$, $z'$, $J$, $H$, and 
$K'$ bands, respectively. 
Exposure times were 180 sec for ROSS2, which acquires simultaneously images in the four Sloan bands, while five ditherings of 7 sec 
each were adopted for each filter of REMIR. 
Details on the reduction of the photometric data are reported in Appendix\,\ref{Appendix:redu_phot}.
The reduction of the X-Shooter spectrum is performed and described in \citet{Manara2016}.

\section{Results}
\label{Sec:Results}

\subsection{Stellar parameters and accretion diagnostics}
\label{Subsec:param}

We analyzed the X-Shooter spectrum with the code \ROTFIT\ \citep{Frasca2017}, which allows us to derive the atmospheric parameters 
(\teff, \logg), the radial velocity (RV), the projected rotational velocity (\vsini), and the veiling ($r$). Details are given in Appendix~\ref{Appendix:analysis}. 
The results of the \ROTFIT\ analysis are summarized in Table~\ref{Tab:param}. The veiling in the red spectral regions analyzed by us is $r<0.2$.   
Our  \teff\ is fully consistent, within the errors, with the value of  3270\,K that was derived by \citet{Manara2016}  from the M4 spectral type (SpT)
and the SpT--\teff\ calibration relation of \citet{Luhman2003}. Moreover, this spectral type corresponds to \teff\,=3200\,K and  \teff\,=3190\,K according to the
SpT--\teff\ relations of \citet{Pecaut2013} and \citet{Herczeg2014}, respectively, which are in perfect agreement with our \teff\ determination.
Both the H$\alpha$ line width at 10\% of the peak, $W_{10\%}=180\pm 18$\,\kms, and the H$\alpha$ flux of 1.2$\times 10^6$\,erg\,cm$^{-2}$s$^{-1}$,
which we measure on the X-Shooter spectrum, indicate ISO-ChaI\,52 as a non-accreting object  (see, e.g., Fig.~11 in \citealt{Frasca2015}) as already noticed 
by \citet{Manara2016,Manara2017}. They observed a small excess in the Balmer continuum (see Fig.~C.1 in \citealt{Manara2016}), which 
translates into an upper limit of $-10.34$  (rescaled to the {\it Gaia} DR2 distance, \citealt{Manara2019}) for $\log \dot{M}_{\rm acc}$ ($M_{\sun}yr^{-1}$), 
and considered it as a doubtful accretor, at a level compatible with the typical chromospheric emission line activity.  
We note that the profiles of the Balmer lines and \ion{Ca}{ii}~K line are all rather narrow and symmetric 
(see Fig.\,\ref{fig:profiles}) with no sign of redshifted absorption components or reversals that are frequently observed in the line profiles of accretors 
\citep[e.g.,][]{Thanathibodee2019,deAlbuquerque2020}. The only notable feature is a wing emission, which is stronger in the red side of the 
H$\beta$ and H$\gamma$ profiles and extends up to $\simeq 200$\,\kms. This is reminiscent of mass flows or turbulence in the upper atmosphere, which are 
mostly observed during flare events \citep[see, e.g.,][]{Doyle1988}. Moreover, as shown in Fig.\,\ref{fig:profiles}, the \ion{He}{i} lines 
$\lambda\lambda 5876, 6876$\,\AA\  are not clearly detected and the \ion{Ca}{ii}\,IRT lines display only a filling in of their cores (Fig.\,\ref{fig:CaIRT}) that 
resembles a purely chromospheric emission. These diagnostics support ISO-ChaI\,52 as a non-accreting or, based on the evidence for UV 
continuum excess derscribed above, a weakly-accreting object. 

\setlength{\tabcolsep}{2pt}

\begin{table}
\caption{Stellar parameters of ISO-ChaI\,52 derived in this work.}	
\begin{tabular}{ccccccc}   
\hline\hline
\noalign{\smallskip}
\teff                   & \logg                  &  RV                          & \vsini                   &  $R_\ast$                      &  $L_\ast$                        & $M_\ast$       \\             
\scriptsize  (K)   & \scriptsize (dex)  & \scriptsize (\kms)  & \scriptsize (\kms)  & \scriptsize ($R_{\sun}$) & \scriptsize ($L_{\sun}$)  & \scriptsize ($M_{\sun}$)  \\
\hline
\noalign{\smallskip}
\scriptsize 3195$\pm$70  & \scriptsize 4.20$\pm$0.35 & \scriptsize 18.1$\pm$3.6 & \scriptsize 13$\pm$6 & \scriptsize 1.14$\pm$0.04 & \scriptsize 0.123$\pm$0.011 & \scriptsize 0.20$\pm$0.05 \\
 \noalign{\smallskip}
 \hline  
\normalsize
\end{tabular}
\label{Tab:param}
\end{table}

\subsection{Spectral energy distribution}
\label{Subsec:SED}

We use the average values of $g'r'i'z'JHK'$ magnitudes outside the dips (Fig.~\ref{fig:multiband}) 
to construct the optical/NIR spectral energy distribution (SED). We extended the SED to the blue side and to mid-infrared (MIR) and far-infrared 
(FIR) wavelengths by adding flux values from the literature. These data are quoted in Table~\ref{Tab:SED}.

\begin{figure}[ht]
\includegraphics[width=8.5cm]{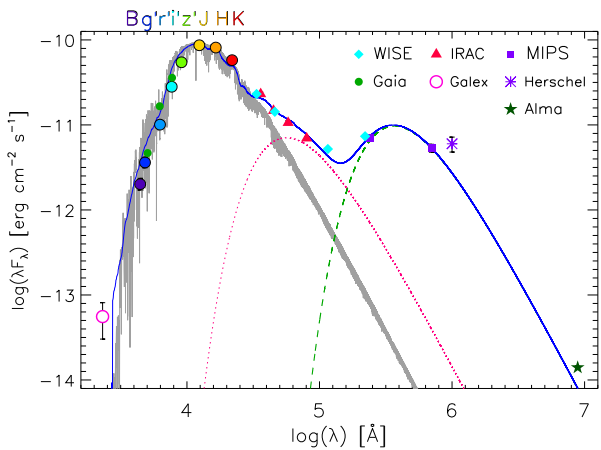}  
\caption{Spectral energy distribution of ISO-ChaI\,52. Mid- and far-infrared fluxes are shown  with different symbols, as indicated in the legend.
The BT-Settl spectrum \citep{Allard2012} that provides the best fit to the star photosphere is shown by a gray line. The two black bodies with $T=650$\,K and 
$T=100$\,K that fit the mid- and far-infrared disk emission are shown by the red-dotted and green-dashed lines, respectively. The continuous blue line displays the sum of the smoothed photospheric 
template and the two black bodies. 
}
\label{fig:SED}
\end{figure}

We adopted the BT-Settl spectrum \citep{Allard2012} with \teff=3200\,K, \feh=0.0, and \logg=4.0, i.e. the one with the parameters closest to those found 
with \ROTFIT, to fit the optical-NIR portion (from $B$ to $J$ band) of the SED (Fig.~\ref{fig:SED}).  Details on the fitting procedure can
be found in Appendix~\ref{Appendix:analysis} and some derived parameters are reported in Table~\ref{Tab:param}. 

The GALEX/NUV flux is clearly in excess with respect to the photosphere. 
NUV flux excess was observed in older M-type stars and was ascribed to stellar magnetic activity \citep[e.g.,][]{Stelzer2013}, but 
it could be also indicative of a mild accretion onto the central star. If this were the case, ISO-ChaI\,52 would be somewhat similar to MY~Lup,
for which accretion is clearly displayed only by UV line and continuum emission revealed by HST \citep{Alcala2019}.   

The SED also displays a significant IR excess at wavelengths longer than about 3\,$\mu$m that is produced by the circumstellar disk. 
The IR excess can be fitted reasonably well with thermal emission from two sources with two different temperatures.  
The MIR emission, which is related to the warmer part of the disk, is fitted with a black body of 650\,K with an emitting area
53 times larger than the stellar surface (red dotted line in Fig.~\ref{fig:SED}), while the FIR emission is reproduced by a source with $T=100$\,K
and an area 1.3$\times 10^5$ times larger that the stellar surface (green dashed line). 
The excess IR luminosity estimated as the sum of these two black-body components, amounts to $L_{\rm disk}\simeq 0.026\,L_{\sun}$, i.e. about
22\% of the stellar luminosity. This is in close agreement with what has been found for accreting objects in Lupus \citep{Merin2008} and 
Cha\,II \citep{Alcala2008}, which all display a fractional disk luminosity $L_{\rm disk}/L_{\ast} > 8\,\%$ that is the limit between passive
reprocessing disks and accretion disks proposed by \citet{Kenyon1987}.
The Hertzsprung-Russell (HR) diagram is shown in Fig.~\ref{fig:HR} along with the pre-main sequence evolutionary tracks and isochrones by \citet{Baraffe2015}.
The position of ISO-ChaI\,52 is between the isochrones at 1 and 3\,Myr and close to the evolutionary track for a 0.2\,$M_{\sun}$ star.

\subsection{REM lightcurves}
\label{Subsec:REM_LC}
\begin{figure}
\begin{center}
\hspace{-0.7cm}
\includegraphics[width=9.8cm]{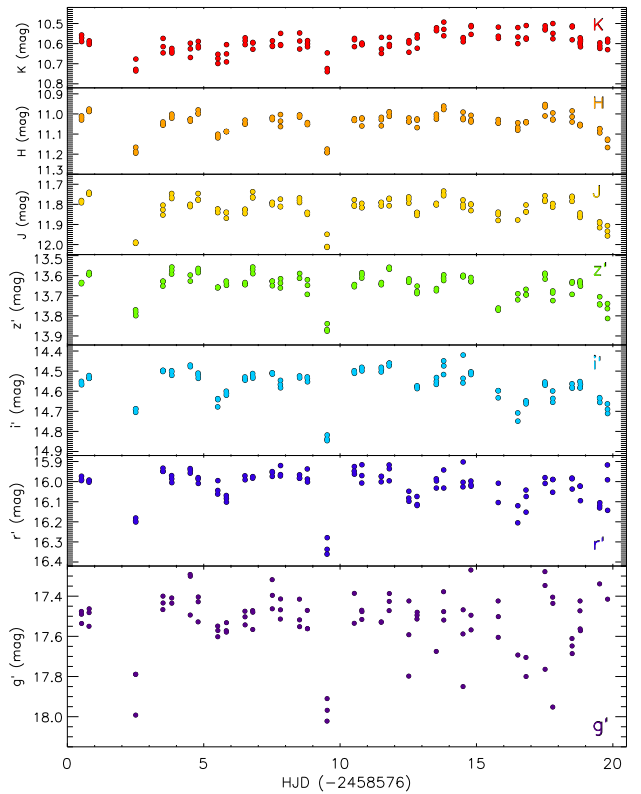}  
\vspace{-0.3cm}
\caption{REM multi-band optical/NIR lightcurves of ISO-ChaI\,52 for the first 20 days of the campaign. 
The scales of the vertical axes have been chosen so as to keep the magnitude ranges constant for a better display 
of the variation amplitudes.}
\label{fig:multiband}
\end{center}
\end{figure}
ISO-ChaI\,52 displays quasi-periodic dimmings throughout the photometric monitoring, which seem to occur about every seven days. 
The light curves observed in the $g'r'i'z'JHK'$ bands are highly correlated and the depth of the dips increases systematically for 
bluer bands (Fig.~\ref{fig:multiband}). These features have been observed in several YSOs both from ground-based and space observations 
\citep[e.g.,][and references therein]{McGinnis2015,Rodriguez2017,Stauffer2017} and have been ascribed to accretion-driven warps in highly inclined 
inner disks, which may be misaligned with respect to the outer disks \citep[e.g.,][]{Bouvier2003,Ansdell2016a, Alencar2018}.	

We searched for the period of these variations by applying a periodogram analysis \citep{sca} to the $r'i'z'JH$ light curves, which are those 
in which the dips are best observed and the photometric errors are low enough. 
To overcome the problem of non-regular data sampling, which introduces aliases in the periodogram, we limited the analysis to the first 
part of the data (50 days for ROSS2 and 20 days for REMIR) and applied the CLEAN iterative deconvolution algorithm \citep{rob}. 
We found for all the bands a peak at about 0.29\,d$^{-1}$, corresponding to a period of about 3.45\,d, with a false-alarm probability $<0.01$,
i.e. with a confidence level $>99$\,\%.  The period uncertainty, evaluated following the prescriptions of \citet{hoba}, is in the 
range 0.01--0.05\,d. 
A string-length analysis \citep{Dworetsky1983} produced similar results, with a first 
deep minimum in string length detected at around 3.5\,d, and subsequent minima of similar or shallower depth at multiples of that period value.
The peak of the periodogram is broader in the NIR band, as expected from the shorter time baseline of REMIR observations (see Fig.~\ref{fig:periodograms}).
We cannot exclude that this period is half of the disk rotation period, because the main dips are about 7 days apart, while smaller dips between them 
are barely visible. The latter ones could be produced by another feature on the opposite side of the disk.
However, a period of $\simeq$\,3.5\,d would hamper the observation of consecutive dips from Earth, due to the day-night cycle. This would explain why
the deeper dimmings are seen at about 7 days from each other in the first part of the data, while they are not clearly visible in the second part 
(Aug-Oct 2019,  see Fig.~\ref{fig:ross2}). The light curve in the $i'$ band folded in phase with the periods of 3.45 and 6.9 days is 
shown in Fig.~\ref{fig:phased_lc}.

\section{Discussion and conclusions}
\label{Sec:discussion}

Assuming that the occulting material is located near the corotation radius, we can take $P_{\rm rot}=3.45$\,d as the star's rotation period 
and derive the inclination of the rotation axis as $\sin i = v\sin i \frac{P_{\rm rot}}{2\pi R_\ast}$,
where we adopted the value of stellar radius given in Table~\ref{Tab:param}, and the more precise value of \vsini\,=\,9.9$\,\pm\,$0.6\,\kms\ reported
by \citet{Nguyen2012}, which was derived from a high-resolution spectrum.
We find $\sin i=0.59\pm0.04$ or $i=36\degr\pm 4\degr$ in which we have considered the errors on \vsini, $R_\ast$, and $P_{\rm rot}$ (0.05\,d). 
This value rules out a nearly edge-on inner disk if our best period, and not twice its value (which we say in Sec.\,\ref{Subsec:REM_LC} we cannot exclude), 
is the real period.  A rotation period of 6.9\,d in the above equation would give $\sin i=1.18\pm0.15$, which is almost exactly $i=90\degr$.
However, an edge-on disk would cause a strong obscuration of the central star, making the object subluminous.
This possibility is ruled out by the position of ISO-ChaI\,52 in the HR diagram (Fig.~\ref{fig:HR}) close to the isochrone at 2\,Myr. 
Therefore, we consider 3.45\,d as the more reliable rotation period. 

The size of the disk can be estimated on the basis of the IR excess. As seen in Sect.~\ref{Subsec:SED}, the FIR part of the SED can be reproduced 
by a region 1.3$\times 10^5$ times larger than the stellar surface emitting as a black-body with \teff=100\,K. 
Under this approximation, such an isothermal disk would have a radius of $R_{\rm disk}\approx$\,3 AU, which would correspond to about 15 mas at the 
distance of ISO-ChaI\,52. This is 30 times smaller than the resolution of the ALMA images collected by \citet{Pascucci2016} in which
the disk is not resolved (see, e.g., their Figs.\,3 and 4).

\begin{figure}
\begin{center}
\hspace{0cm}
\includegraphics[width=8.5cm]{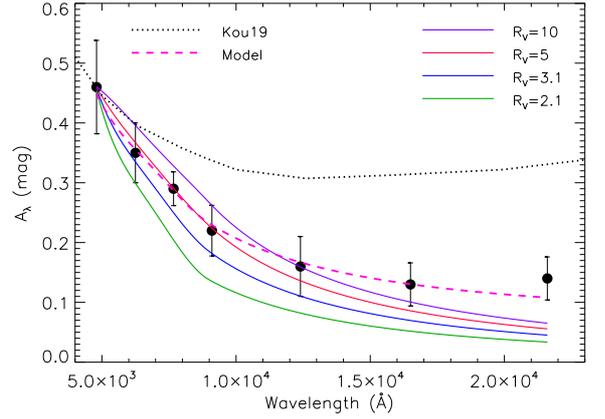}  
\vspace{-0.2cm}
\caption{Extinction taken as the amplitude of the second dip (observed at JD=2458585) as a function of the filter wavelength (dots).
The full lines represent extinction laws at different values of $R_V=A_V/E(B-V)$ according to \citet{Cardelli1989}.
The black dotted line is the model by \citet{Koutoulaki2019}. The dashed line represents our two-component model
(Eq.~\ref{Eq:model}).
All curves are normalized to the observed extinction in the $g'$ band ($\lambda_c=4800$\,\AA). 
}
\label{fig:dip_ampl}
\end{center}
\end{figure}

The depth of the dips in bands of different wavelengths provides useful information on the material occulting
the central star. In particular, in the optical bands, the luminosity dimming is largely due to the dust, which affects
in a different way bluer and redder bands depending on the average size of the grains.
We display the extinction $A_\lambda$, taken as the depth of the best observed dip (JD=2458585), as a function of wavelength in Fig.~\ref{fig:dip_ampl}.
Bearing in mind the radiative transfer equation for a purely absorbing medium,  $I=I_0e^{-\tau_{\lambda}}$, the extinction $A_\lambda$ is proportional to 
the effective optical depth $\tau_{\lambda}$ of the layer ($A_{\lambda}=1.086\,\tau_{\lambda}$). 
We have compared the extinction observed during the dip with the extinction law by \citet{Cardelli1989}.
It is evident from Fig.~\ref{fig:dip_ampl} that the observed extinction is flatter than the one typical of the interstellar medium (ISM), which has a 
total-to-selective extinction ratio $R_V=A_V/E(B-V)=3.1$. It is best reproduced in the $g'r'i'z'$
bands by a Cardelli law with $A_V=0.41$\,mag and $R_V=5.0$, which implies an average grain size larger than in the ISM.
However, no value of $R_V$ is able to reproduce the extinction up to the NIR.

A slope in the extinction law flatter than that of the ISM and sometimes a nearly gray extinction have been reported for AA~Tau, 
the prototype of dippers, by \citet{Bouvier1999,Bouvier2003} and for the dips of  LkCa\,15 by \citet{Alencar2018}.
\citet{Koutoulaki2019} were able to reproduce the extinction curve observed during a dimming event of RW~Aur observed with X-Shooter,
which is much flatter than the ISM, with a power-law distribution of grain size from a minimum value $a_{\rm min}=0.1\,\mu$m to a maximum 
$a_{\rm max}=150\,\mu$m, including the  scattering in the effective optical depth.  Neglecting the scattering would require even larger grains. 
An ISM-type extinction ($R_V=3.1$) would be instead produced by much smaller grains ($a_{\rm max}=0.1\,\mu$m). 
In Fig.~\ref{fig:dip_ampl}, we have overplotted the extinction model by \citet{Koutoulaki2019}, scaled to the observed extinction in the $g'$ band.
We note that this curve is flatter than our data, suggesting different conditions for the region causing the dips in ISO-ChaI\,52, as also expected	
from the much deeper dips of RW~Aur (2--3 mag in the $V$ band) compared to  0.3-mag dips in $r'$ for ISO-ChaI\,52. 
Another dipper that displays an extinction effect similar to ISO-ChaI\,52 is  V354~Mon \citep{Fonseca2014}. \citet{Schneider2018} used X-Shooter spectra of this 
star taken outside and within the dips to derive the properties of the obscuring dust in the disk warp. They showed that it is not possible to reproduce the entire
dimmed spectrum  by applying a single reddening law to the uneclipsed spectrum. The blue-visible part of the dimmed spectrum could be roughly reproduced 
by applying a Cardelli law with $A_V\simeq 1.2$\,mag and $R_V=6.0$ to the uneclipsed spectrum, but this left an IR flux excess, similar to what we find for 
ISO-ChaI\,52. 
A more evolved model, which includes an upper disk layer with an ISM-type extinction and  an opaque disk region producing a gray extinction, was able to 
reproduce their data.   
We have applied to our data a simple two-component model similar to that of \citet{Schneider2018}. In this case, the extinction of the stellar light can be
expressed as
\begin{equation}
I=I_0(\alpha\,e^{-\tau_{\rm G}}\, +\, \beta\,e^{-\tau_{\lambda}}), 
\label{Eq:model}
\end{equation}
\noindent{where $\alpha$ and $\beta$ are the fractions of the stellar disk occulted by the opaque-gray and thin region of the circumstellar disk, respectively,
and $\tau_{\rm G}$ the optical depth of the gray layer.	
The best-fitting model (dashed line in Fig.~\ref{fig:dip_ampl}) has $\alpha=0.38$, $\beta=0.62$,  $\tau_{\rm G}=0.22$ ($A_{\rm G}\simeq0.24$\,mag), and 
$\tau_{\lambda=5500}\simeq 0.51$ ($A_{V}=0.55$\,mag) with $R_V=3.1$.
Although this is a simple schematic model, it tells us that the feature occulting the central star must contain both small-sized grains ($a<0.5\,\mu$m),
producing an ISM-type extinction, and larger ones giving rise to a much flatter or gray extinction.

The fact that we observe dips in a weakly-accreting object rises the question on the connection between the strength of accretion and the presence of a warp in 
YSO disks. \citet{Stauffer2015} have analyzed a specific category of dipper stars, which share with ISO-ChaI\,52 the late SpT and the short-duration 
($\simeq$~1\,d) and shallow (0.1--0.4 mag) dips. They explore alternative scenarios to warped inner disks to explain the origin of the dips, 
 including the occultation of the star by spiral-arm overdensities in the inner disk raised by an embedded planet or by dust entrained into 
an accretion funnel. The latter model can also explain dips for low-inclination objects. Dippers with nearly face-on outer disks have already been found 
\citep[e.g.,][]{Ansdell2016a,Scaringi2016}. In particular,  \citet{Ansdell2020} found the disk inclination distribution to be consistent with isotropic. 
The possible explanations they propose include dust clouds driven by disk winds (which can determine dips in systems with inclinations as low as $\sim$\,30\degr), or misalignments between inner disk and outer disk (which might be caused by a substellar or planetary companion).
Transits of cometary-mass objects have been also proposed for a few non-accreting dippers \citep[e.g.][]{Scaringi2016,Ansdell2019}.
A dipping behavior may also be observed in objects seen at mid-inclinations when the dipole magnetic field of the star exhibits a small tilt angle with respect to its 
rotation axis, which leads to the formation of accretion streams that extend high above the disk midplane \citep{Bodman2017}.
In the following we consider the conditions under which accretion-driven structures, such as disk warps or funnels, can produce dips in ISO-ChaI\,52. 

\citet{Bessolaz2008} have studied the conditions for a steady accretion flow from a circumstellar disk in the presence of a dipolar stellar 
magnetic field. There are few measures of photospheric magnetic fields for low-mass ($M_\ast\leq 0.5\,M_{\sun}$) YSOs with typical 
values of $B_\ast\approx$\,1\,kG \citep[e.g.,][]{Hill2019,Lavail2019}, but fields on the order of 100\,G and lower have been also observed 
\citep[e.g.][]{Donati2010,Morin2011}.
From Eq.~6 of \citet{Bessolaz2008} and adopting the stellar parameters in Table~\ref{Tab:param} and the upper limit 
$\dot{M}_{\rm acc}\leq 4.6\times 10^{-11}\,M_{\sun}yr^{-1}$ \citep{Manara2019}, we derive for ISO-ChaI\,52 a disk truncation radius $R_{\rm T}\simeq 25 R_{\sun}$ 
for a magnetic field $B_{\ast}=1$\,kG.
For comparison, the Keplerian corotation radius for $P_{\rm rot}=3.45$\,d, i.e. the period we observe for the dips, is $R_{\rm C}\simeq 5.6 R_{\sun}$, which would be 
8.9\,$R_{\sun}$ adopting $P_{\rm rot}=6.9$\,d.
The case of $R_{\rm T} > R_{\rm C}$ corresponds to a propeller regime of star-disk interaction \citep{Ustyugova2006}, where stable funnel-flow accretion is 
inhibited. 
Therefore, to have a steady accretion regime, the photospheric magnetic field should be lower: $B_{\ast}\leq 70$\,G for $P_{\rm rot}=3.45$\,d and  $B_{\ast}\leq 150$\,G if the period is twice. Another possible explanation is that the magnetic field is strong, but we are underestimating the mass accretion rate. 
A truncation radius $R_{\rm T}\leq 8.9 R_{\sun}$ with a field $B_\ast=1$\,kG would require a mass accretion rate $\dot{M}_{\rm acc}\ge 3\times 10^{-9}\,M_{\sun}yr^{-1}$, 
i.e. about two orders of magnitude larger than the observed upper limit; a larger mass accretion rate is needed for $R_{\rm T}\leq 5.6 R_{\sun}$. 
Indeed, there are examples of YSOs with apparently low or no accretion from optical tracers, but with significant accretion as drawn from near-UV and far-UV 
observations \citep[e.g.,][]{Alcala2019}. We note, however, that these objects are hotter than ISO-ChaI\,52, hampering the detection of Balmer 
continuum excess emission due to a low contrast with respect to the photospheric emission. Should $\dot{M}_{\rm acc}$ in ISO-ChaI\,52 be so high, 
a much stronger UV excess  and Balmer continuum than observed would have been detected. 

 To conclude, we think that a low-accretion rate coupled with a relatively weak surface magnetic field can give rise to disk warps or accretion 
structures able to produce dips in this low-mass YSO. This work shows the effectiveness of long-term simultaneous multiband photometry ranging from the optical 
to the NIR domain for the study of the circumstellar environment in YSOs.  

\begin{acknowledgements}
We thank the anonymous referee for her/his useful comments and suggestions.
We acknowledge the support from the Italian {\it Ministero dell'Istruzione, Universit\`a e  Ricerca} (MIUR).
This work has been partially  supported by the project  PRIN-INAF-MAIN-STREAM 2017
``Protoplanetary disks seen through the eyes of new-generation instruments''. 
CFM acknowledges an ESO fellowship. This project has received funding from the European Union's Horizon 2020 
research and innovation programme under the Marie Sklodowska-Curie grant agreement No 823823 (DUSTBUSTERS).
This work was partly supported by the Deutsche Forschungs-Gemeinschaft (DFG, German Research Foundation) - 
Ref no. FOR 2634/1 TE 1024/1-1.
LV acknowledges support by an appointment to the NASA Postdoctoral Program at the NASA Ames Research Center, 
administered by Universities Space Research Association under contract with NASA.
GR acknowledges funding from the Dutch Research Council (NWO) with project number 016.Veni.192.233.
This research made use of SIMBAD and VIZIER databases, operated at the CDS, Strasbourg, France. 

\end{acknowledgements}

\bibliographystyle{aa}

{}

\newpage

\appendix

\section{Photometric data reduction}
\label{Appendix:redu_phot}

 \begin{figure} 
 \hspace{0.5cm}
     \includegraphics[width=12cm]{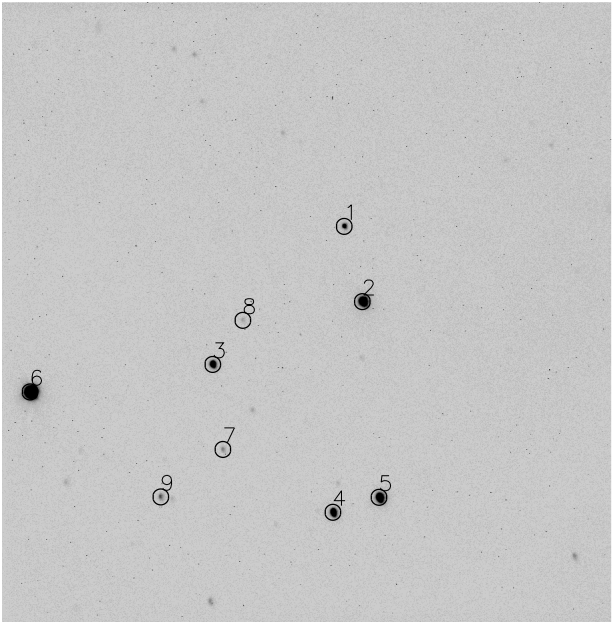}  
  \caption{Field of ISO-ChaI\,52 as observed by ROSS2 camera with the $i'$ filter. The identification code (ID in 
	   Table~\ref{Tab:sources}) is written next to the stars for which we extracted magnitudes. ID\,=\,1 for  ISO-ChaI\,52. }
  \label{fig:field}
  \end{figure}

For the ROSS2 camera, we have generated master flats using the twilight flat-fields taken during the observing run, which are 
available in the REM archive. The latter were used to correct for pixel-to-pixel sensitivity variations, as well as 
for the vignetting and illumination of the field of view.
Each scientific image, after subtraction of the dark-frame, was divided by the proper master-flat, depending on the
filter. 

The field of view, as observed in the $i'$ filter, along with the identification code for our target and the comparison stars reported in Table~\ref{Tab:sources}, 
is displayed in Fig.~\ref{fig:field}. 

\setlength{\tabcolsep}{4pt}

\begin{table*}
\caption{Literature data for some stars in the field of ISO-ChaI\,52.}
\begin{center}
\begin{tabular}{lcccccccccc}   
\hline\hline
\noalign{\smallskip}
ID$^{a}$ & Name  & 2MASS &  $g'$ &  $r'$  &  $i'$  &  $z$   &  $J$  &  $H$  &  $K'$      & $\pi^{b}$    \\             
         &       &       & (mag) &  (mag) &  (mag) &  (mag) &  (mag) &  (mag) &  (mag) &   (mas)       \\
\hline
\noalign{\smallskip}
 1  & \object{ISO-ChaI\,52}    & J11044258-7741571  & \dots    &  \dots   &  \dots   &  13.549  &  11.814  &   11.002  &   10.642  & 5.18$\pm$0.07 \\
 2  &                          & J11042217-7741319  &  13.377  &  12.187  &  11.702  &  11.245  &   9.738  &	9.167  &    8.913  & 5.20$\pm$0.02 \\ 
 3  & \object{ISO-ChaI\,35}    & J11035902-7743349  &  15.792  &  14.224  &  13.296  &  12.349  &  10.323  &	9.479  &    9.050  & 3.16$\pm$0.04  \\ 
 4  & \object{Glass~L}         & J11032288-7741301  &  16.587  &  14.297  &  13.140  &  12.155  &   9.926  &	8.691  &    8.264  & 0.40$\pm$0.04 \\ 
 5  & \object{Glass~M}         & J11032892-7740518  &  16.180  &  13.437  &  12.058  &  10.934  &   8.416  &	7.044  &    6.523  & 0.75$\pm$0.05 \\ 
 6  & \object{Tyc\,9414-768-1} & J11034449-7746111  &  11.385  &  10.650  &  10.626  &  10.503  &   9.547  &	9.039  &    8.934  & 9.97$\pm$0.02 \\ 
 7  &                          & J11033587-7743146  & \dots    &  \dots   &  \dots   &  14.984  &  12.668  &   11.415  &   11.024  & 0.49$\pm$0.08  \\  
 8  &                          & J11041245-7743144  & \dots    &  \dots   &  \dots   &  15.625  &  13.240  &   12.037  &   11.525  & 2.31$\pm$0.10  \\ 
 9  &                          & J11032037-7744028  & \dots    &  \dots   &  \dots   &  14.572  &  12.589  &   11.728  &   11.351  & 1.88$\pm$0.05 \\ 
 \noalign{\smallskip}
\hline  
 \noalign{\smallskip}
\end{tabular}
\end{center}
~\\ \textbf{Notes} $^a$ Identification code as in Fig.~\ref{fig:field}. $g'r'i'$ magnitudes from APASS \citep{APASS}.\\
 $z$ magnitudes from SkyMapper \citep{SkyMapper}. $JHK'$ magnitudes from 2MASS \citep{2MASS,skrutskieetal2006}.\\
 $^b$ Parallax from {\it Gaia} DR2 \citep{GaiaDR2}.
\label{Tab:sources}
\end{table*}

The pre-reduction of the REMIR images is automatically done by the AQuA pipeline \citep{Testa04} and the co-added and 
sky-subtracted frames, resulting from five individual ditherings, are made available to the observer. 

Aperture photometry for all the stars listed in Table~\ref{Tab:sources} was performed with DAOPHOT by using the IDL\footnote{IDL 
(Interactive Data Language) is a registered trademark of  Harris Corporation.} routine \textsc{Aper}.
The photometric errors based on the photon statistics in the NIR bands are typically in the range 0.008--0.018 mag for ISO-ChaI\,52 
($H\simeq 11\fm 0$) with average values of 0.013, 0.008, and 0.015 mag in $J$, $H$, and $K'$, respectively.
They range instead from 0.003 to 0.006 mag for a brighter star like ISO-ChaI\,35 ($H\simeq 9\fm 5$).
In the optical bands the average photometric errors for ISO-ChaI\,52 are of 0.013, 0.008, 0.020, and 0.045 mag for $z'$, $i'$, $r'$, and 
$g'$, respectively.

As a result of the field rotation, the center of the field can vary in different pointings of the telescope by as much as a few arcmin, 
so that only three stars (\#1, \#2, and \#3) are included in all the useful images.
We have therefore chosen \#2=\object{2MASS\,J11042217-7741319}, which is the brightest among these three stars and the closest to 
ISO-ChaI\,52, as comparison object for the purpose of differential photometry.
This object is not reported in the literature as a Cha\,I member, even if its parallax $\pi=5.2009$\,mas is nearly the same as that of ISO-ChaI\,52.
For each band, we have added the magnitude of \#2 listed in Table~\ref{Tab:sources} to get the magnitude of ISO-ChaI\,52  and the other stars in the field.
The latter stars, even if with less data points, allowed us to check for any eventual variability of \#2 and to evaluate the {\it final} 
data uncertainty as the rms of their magnitude differences. 
To this aim, we have calculated the magnitude of \#4=\object{Glass~L} using as comparison \#5=\object{Glass~M}, which are close to each other
and both background stars unrelated to the Cha\,I cloud. The rms scatter of the photometry of Glass~L all along the observing season is 0$\fm$021, 
0$\fm$022, 0$\fm$024, 0$\fm$047, for the $z'$, $i'$, $r'$, and $g'$ band, respectively.

As an example we show in Fig.\,\ref{fig:ross2} the light curves of ISO-ChaI\,52, ISO-ChaI\,35 (\# 3) and Glass\,L (\#4).
The points of ISO-ChaI\,35 at JD\,$\simeq 2458583.8$, which are about 0.15\,mag brighter than those taken 7 hours before and 17 hours later, 
are likely taken during a flare, because a similar event is also observed in the $g'$ band as an enhancement of about 0.35 mag, while it is not
visible in the redder bands.  
If we consider the magnitude difference between \#2 and \#4, the rms increases slightly to just 0.03\,mag and no clear periodicity appears 
in the data, which reassures us on the use of \# 2 as a comparison for ISO-ChaI\,52.

\begin{figure}
\begin{center}
\hspace{0cm}
\includegraphics[width=9.5cm]{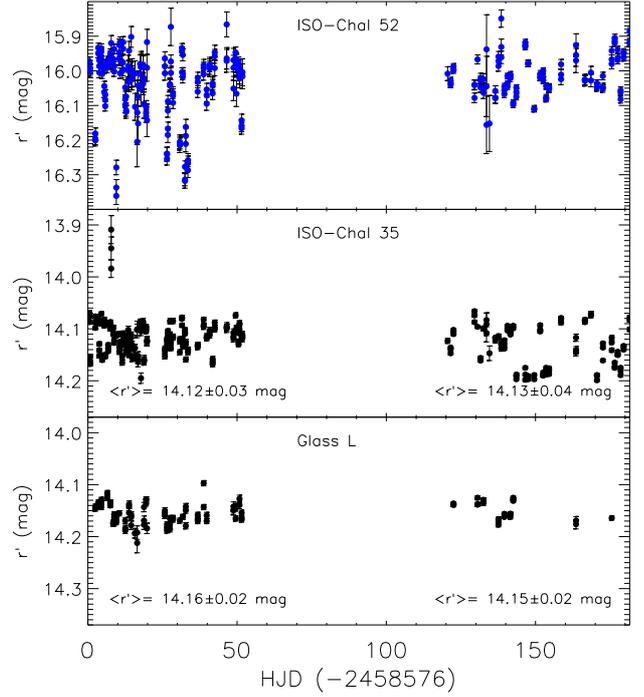}  
\vspace{-0.2cm}
\caption{ROSS2 $r'$ lightcurve of ISO-ChaI\,52 (top panel), ISO-ChaI\,35 (middle panel) and Glass\,L (bottom panel).
We have used \object{2MASS\,J11042217-7741319} (\#2) for the first two stars and \object{Glass~M} (\#5) for the latter, as
comparison stars, adopting the magnitudes listed in Table~\ref{Tab:sources}.
The average magnitudes of the last two stars and their rms scatter are provided in the respective panels for the two data segments.
}
\label{fig:ross2}
\end{center}
\end{figure}

\section{ROTFIT and SED analysis}	
\label{Appendix:analysis}
The code \ROTFIT\  finds the best photospheric template spectrum \citep[here: BT-Settl,][]{Allard2012} that reproduces the target spectrum by minimizing 
the $\chi^2$ of the difference between the observed and synthetic spectra in specific spectral segments. The spectral intervals selected for the analysis with 
\ROTFIT\ are normalized to the local continuum and contain features that are sensitive to the effective temperature and/or surface gravity, such as the 
\ion{Na}{i}  doublet at $\lambda\approx$\,819 nm and the \ion{K}{ i} doublet at  $\lambda\approx$\,766--770 nm (see Fig.\,\ref{fig:ROTFIT}).

\begin{figure}
\includegraphics[width=5.5cm]{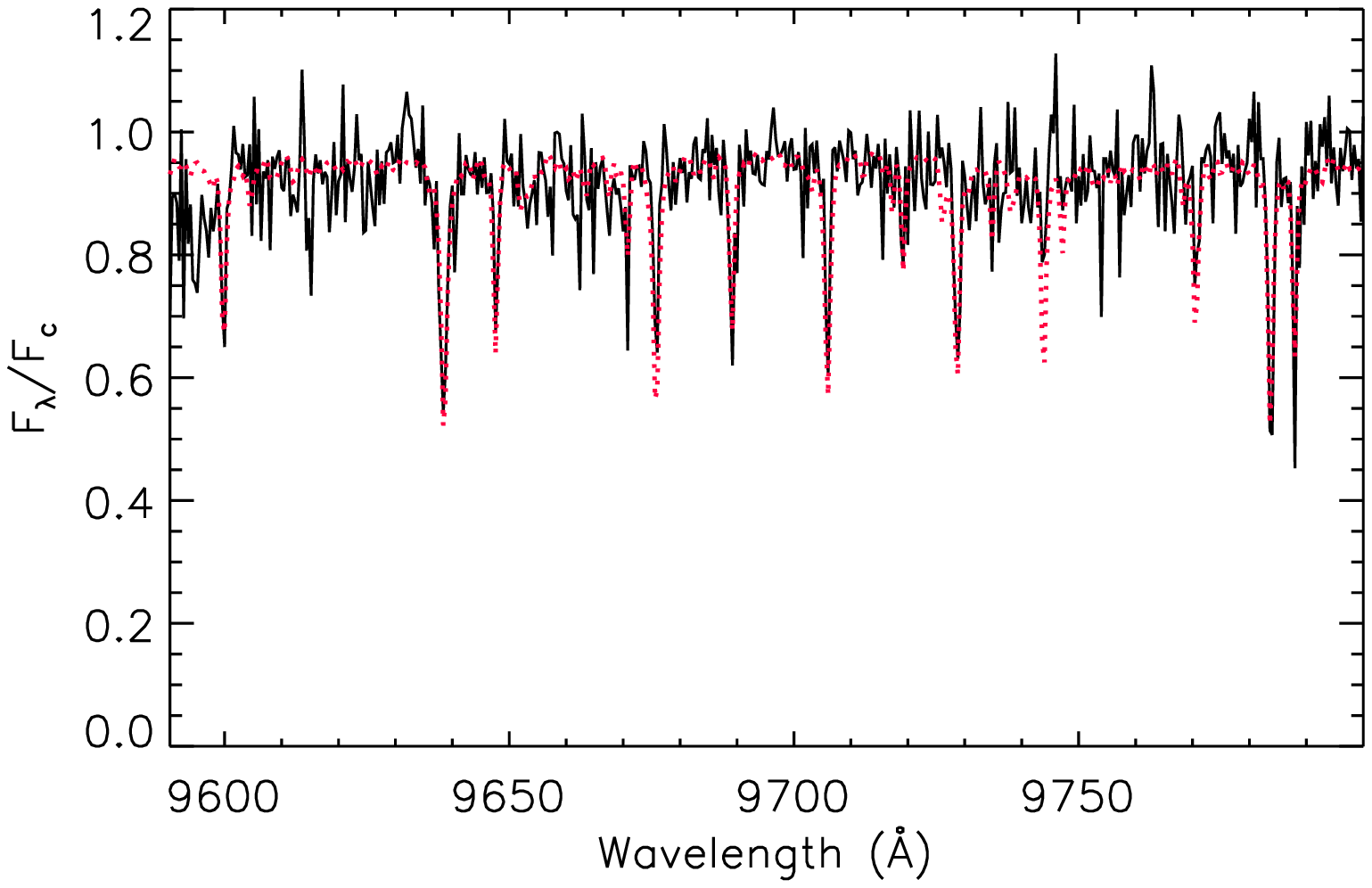}   
\hspace{-.5cm}
\includegraphics[width=3.4cm]{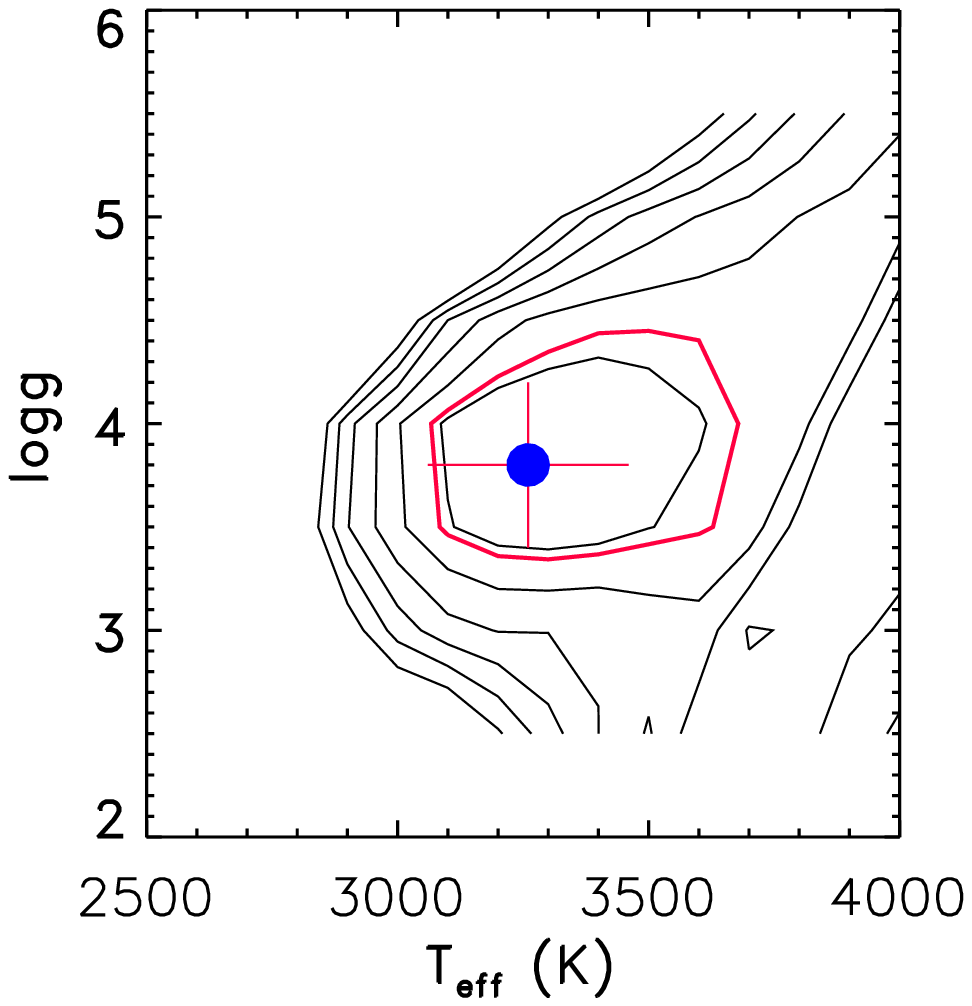}   
\includegraphics[width=5.5cm]{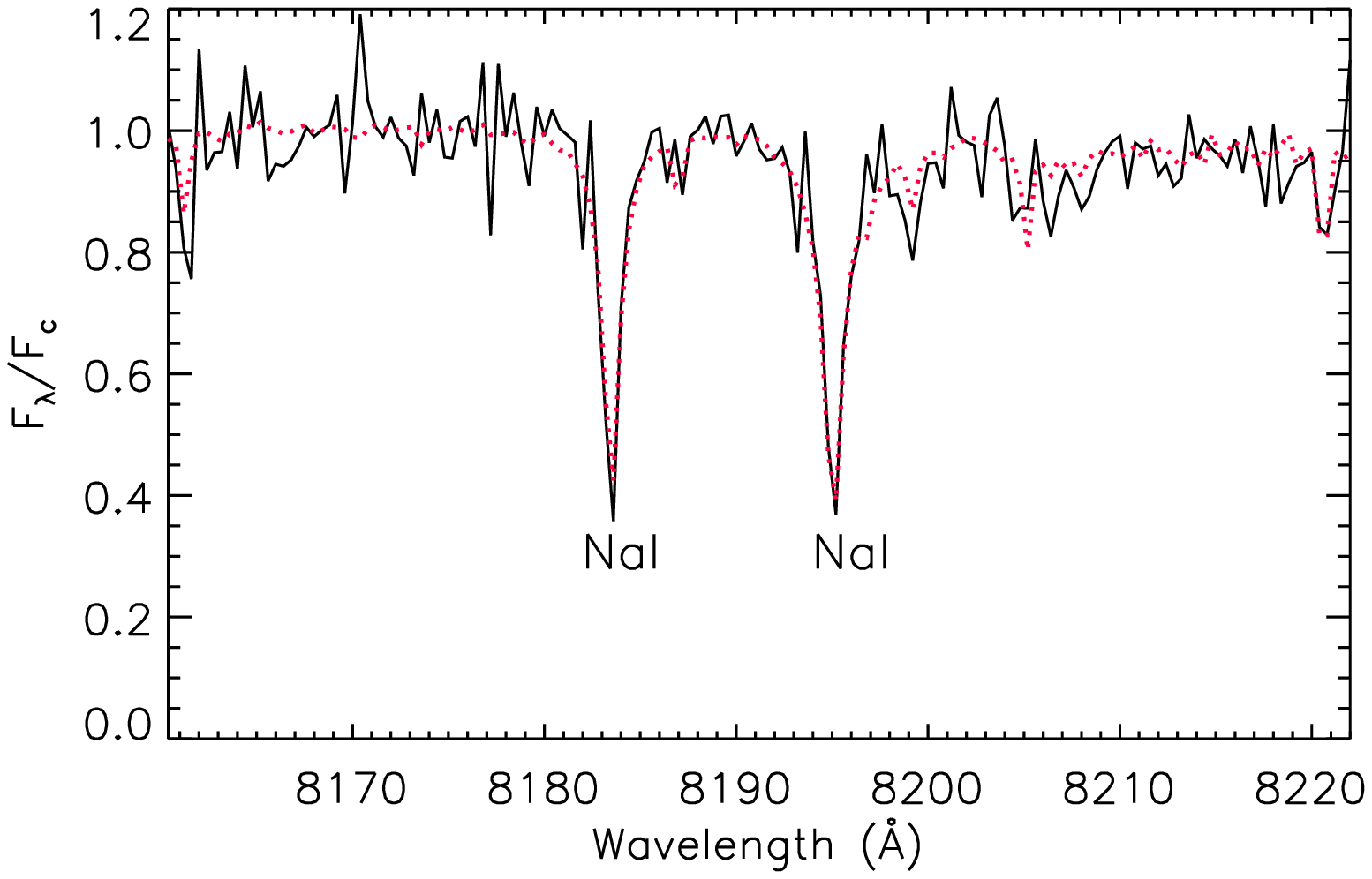}  
\includegraphics[width=3.4cm]{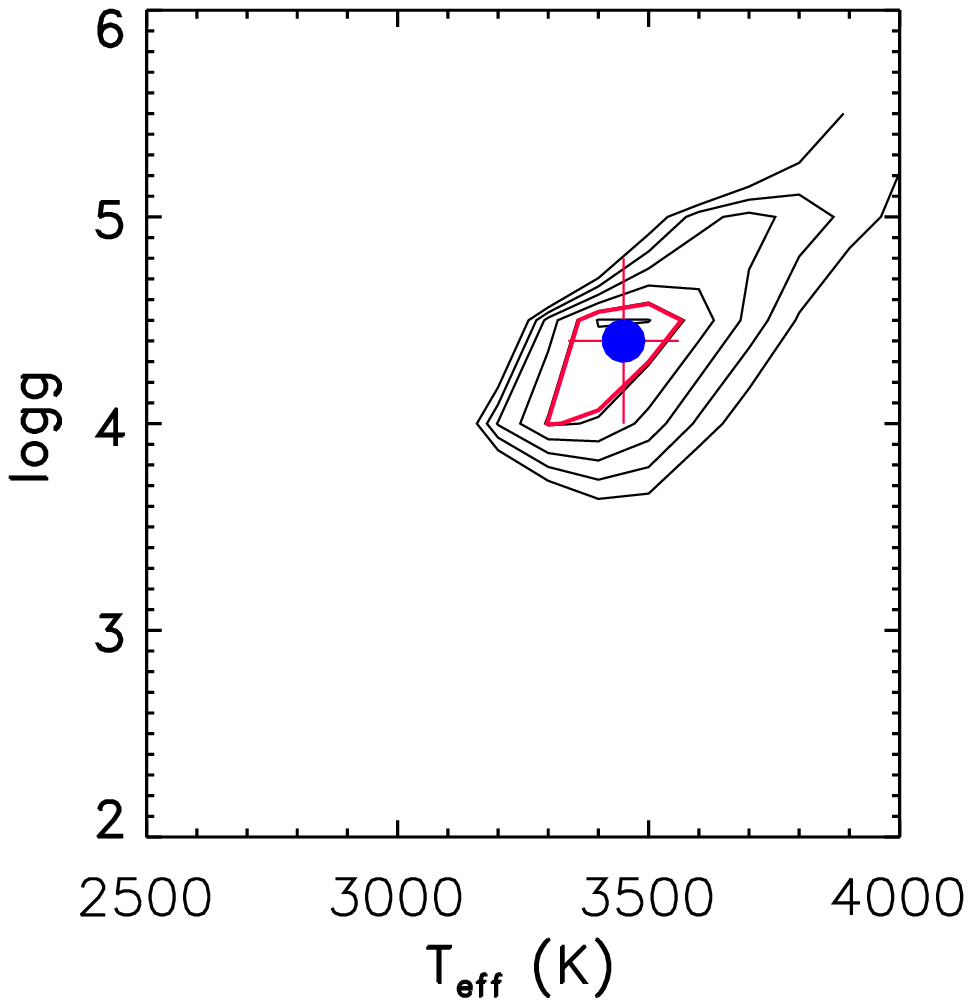}  
\includegraphics[width=5.5cm]{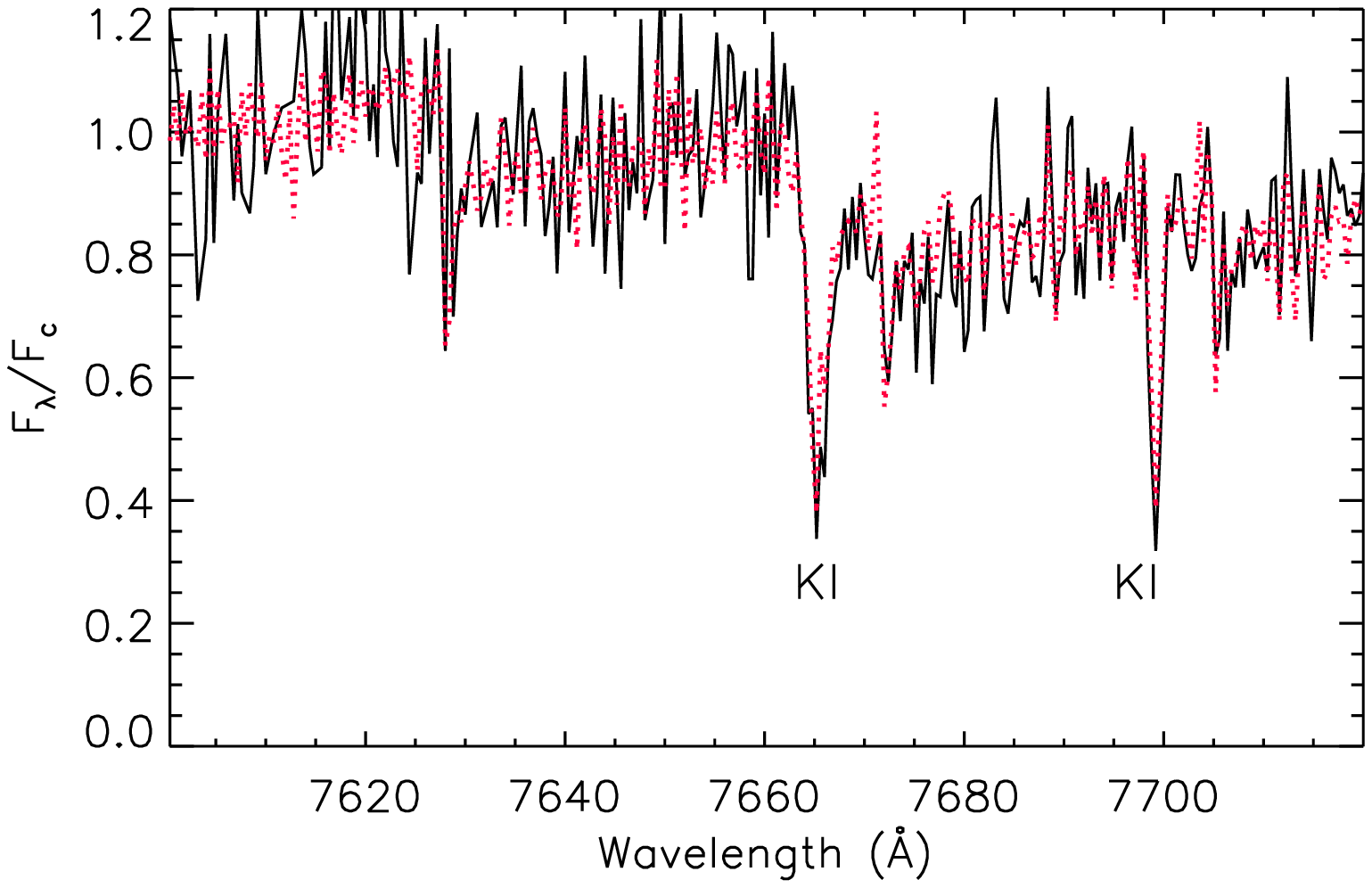}  
\includegraphics[width=3.4cm]{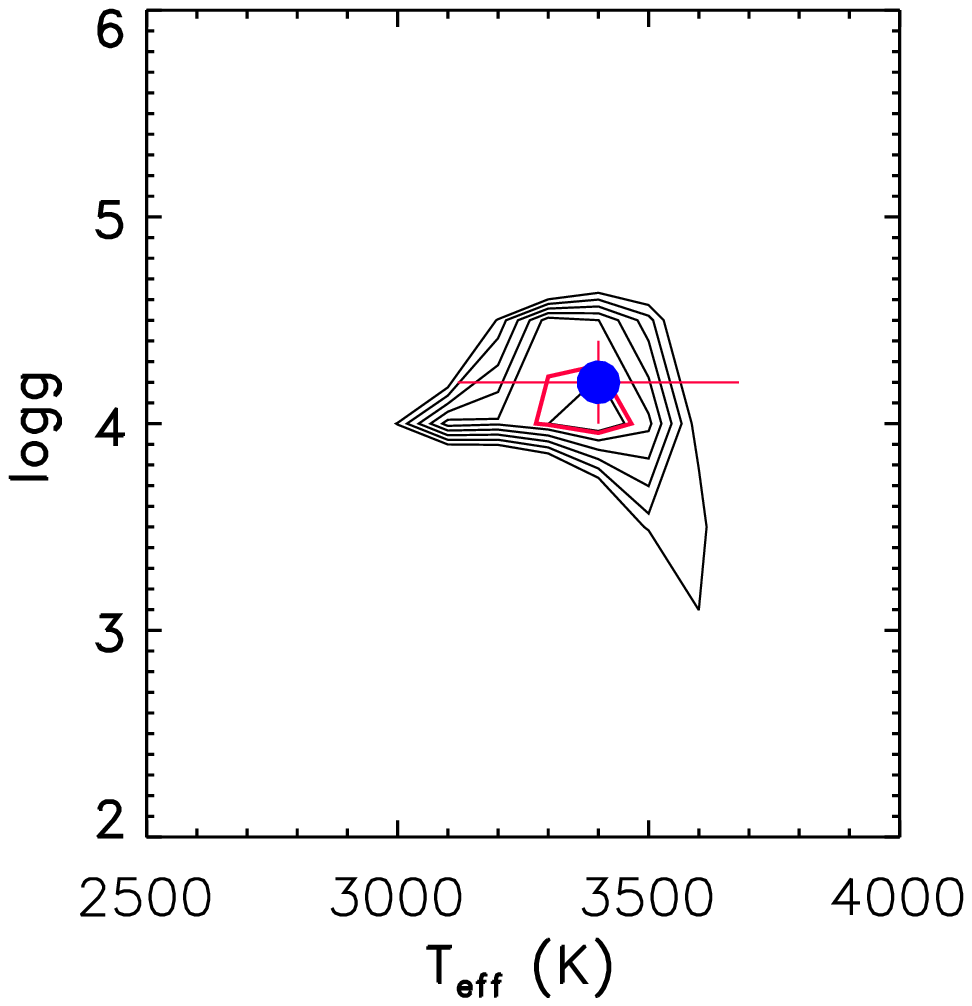}  
\vspace{-.3cm}
\caption{{\it Left panels:} continuum-normalized VIS X-Shooter spectrum of ISO-ChaI\,52 in three regions (black full lines) with the best fitting synthetic spectrum  
overplotted (red dotted lines). {\it Right panels:} $\chi^2$ contour maps in the \teff-\logg\ plane. In each panel, the 1$\sigma$ confidence level is 
denoted by the red contour. The best values and errorbars on \teff\ and \logg\ are also indicated. 
}
\label{fig:ROTFIT}
\end{figure}

The SED was built by complementing the $g'r'i'z'JHK'$ magnitudes observed with REM outside the dips with literature values. Notably, we have 
added the Johnson $B$ magnitude reported in the TIC (TESS input catalog, \citealt{TESS}) and the GALEX-DR5 NUV flux of 4.28\,$\mu$Jy at 2316\,\AA\ \citep{Bianchi2011}
at shorter wavelengths.
The mid-infrared (MIR) and far-infrared (FIR) fluxes were retrieved from the WISE data release \citep{WISE}, from {\it Spitzer} IRAC and MIPS data 
\citep{Dunham2015}, and Herschel/PACS 100\,$\mu$m \citep{Ribas2017}.
We have also included the sub-mm flux at $\lambda$=887\,$\mu$m, $F_{\nu}=4.15\pm0.16$\,mJy, reported by \citet{Pascucci2016}, which is a 
disk-integrated value, as the disk is not resolved in the ALMA image (see, e.g., Figs. 3 and 4 in \citealt{Pascucci2016}).
All these values are reported in Table~\ref{Tab:SED}.

The distance $d=193\,\pm\,3$\,pc  was calculated from the {\it Gaia}\,DR2 parallax of 
ISO\,ChaI\,52 ($\pi=5.18\pm 0.07$\,mas) as $d=1000/\pi$. This large value of parallax allows us to neglect small 
corrections like those proposed by \citet{Lindegren2018} that would decrease the distance by only 1\,pc, which is less than the distance error.
In the fitting procedure, applied to the fluxes from $B$ to $J$ band, we fixed the distance and the effective temperature and let the stellar radius, 
$R_{\ast}$, and the extinction, $A_V$, vary until a minimum $\chi^2$ was reached. 
The key parameter affecting the results is the effective temperature, therefore we run the code also fixing \teff\ to the extreme 
values given by the  \teff\ error of 70\,K.		
We found $A_V=0.43\pm0.32$\,mag, which is lower than the value of 1.2\,mag reported by \citet{Manara2016}, who analyzed the full
calibrated X-Shooter spectrum and used real-star spectra of slighlty higher \teff\  as templates. We note that $A_V$ is very sensitive to the intrinsic shape of 
the flux distribution, i.e. to \teff, while the stellar radius, $R_{\ast}=1.14\pm0.04\,R_{\sun}$, is strongly dependent on the distance
and on the observed flux near the maximum of the SED, where the extinction plays a minor role.

The stellar luminosity, calculated as $L_{\ast}=4\pi R_\ast^2\sigma T_{\rm eff}^4$, is $L_{\ast}=0.123\pm 0.011\,L_{\sun}$, where the error
takes into account the $R_\ast$ and \teff\ errors and their covariance.  This value of $L_{\ast}$ is higher than 
the luminosity of 0.09\,$L_{\sun}$ reported by \citet{Manara2016}, who adopted a distance $d=160$\,pc. However, the latter becomes 0.13\,$L_{\sun}$ with
the {\it Gaia} distance $d=193$\,pc, in good agreement with our determination.

\setlength{\tabcolsep}{3pt}

\begin{table}
\caption{Data for the SED of ISO-ChaI\,52.}	
\begin{tabular}{lcccl}   
\hline\hline
\noalign{\smallskip}
Band      & $\lambda_{\rm c}$ &  Magnitude & Flux                                                    &   Reference   \\             
             & ($\mu$m)               & (mag)          & \scriptsize{(erg\,cm$^{-2}$s$^{-1}$\AA$^{-1}$)} & \\
\hline
\noalign{\smallskip}
$NUV$ &        0.231             &     22.32$\pm$0.48          &   (2.41$\pm$1.10)E-17      & B2011 \\ 
 $B$     &        0.444             &     18.57$\pm$0.16          &   (2.68$\pm$0.40)E-16      & S2019 \\ 
 $g'$    &        0.485             &     17.50$\pm$0.15              &   (4.68$\pm$0.63)E-16      &   \scriptsize{Present work}   \\
 $r'$     &        0.621             &     15.95$\pm$0.10              &   (1.16$\pm$0.11)E-15      &   \scriptsize{Present work}   \\
 $i'$     &        0.767             &     14.52$\pm$0.07              &   (2.88$\pm$0.19)E-15      &   \scriptsize{Present work}   \\
 $z'$    &        0.910             &     13.55$\pm$0.07              &   (5.02$\pm$0.32)E-15      &   \scriptsize{Present work}   \\
$BP$    &        0.505            &      17.146$\pm$0.014          &   (5.77$\pm$0.08)E-16      & {\it Gaia}\,DR2 \\ 
$G$     &        0.623            &      15.218$\pm$0.003          &   (2.09$\pm$0.01)E-15      & {\it Gaia}\,DR2 \\ 
$RP$   &         0.772            &      13.892$\pm$0.008 	     &   (3.65$\pm$0.03)E-15       & {\it Gaia}\,DR2 \\ 
 $J$      &        1.24              &     11.75$\pm$0.07              &   (6.24$\pm$0.42)E-15      &   \scriptsize{Present work}   \\
 $H$     &        1.65              &     10.98$\pm$0.06              &   (4.59$\pm$0.25)E-15      &   \scriptsize{Present work}   \\
 $K'$     &        2.19              &     10.58$\pm$0.06             &   (2.51$\pm$0.14)E-15      &   \scriptsize{Present work}   \\
$WISE\,1$  &   3.35              &     10.186$\pm$0.023        &   (6.89$\pm$0.15)E-16       &  W2010 \\ 
$WISE\,2$  &   4.60              &      ~9.728$\pm$0.020         &   (3.10$\pm$0.06)E-16       &  W2010 \\ 
$WISE\,3$  & 11.56              &      ~7.901$\pm$0.019         &   (4.50$\pm$0.08)E-17       &  W2010 \\ 
$WISE\,4$  & 22.09              &      ~5.461$\pm$0.030         &   (3.33$\pm$0.09)E-17       &  W2010 \\ 
$IRAC\,1$  &   3.6               &     \dots          &   (6.42$\pm$0.33)E-16  &   D2015 \\ 
$IRAC\,2$  &   4.5               &     \dots          &   (3.27$\pm$0.15)E-16  &   D2015 \\ 
$IRAC\,3$  &   5.8               &     \dots          &   (1.83$\pm$0.09)E-16  &   D2015 \\ 
$IRAC\,4$  &   8.0               &     \dots          &   (8.71$\pm$0.41)E-17  &   D2015 \\ 
$MIPS\,24$  &  24              &     \dots           &   (2.95$\pm$0.11)E-17  &   D2015 \\ 
$MIPS\,70$  &  70              &     \dots           &   (7.64$\pm$0.82)E-18  &   D2015 \\ 
$Herschel$  &  100            &     \dots           &   (6.00$\pm$1.20)E-18  &   R2017 \\ 
$ALMA$      &   887            &     \dots           &   (1.58$\pm$0.06)E-21  &   P2016 \\ 
 \noalign{\smallskip}
 \hline  
\normalsize
\end{tabular}
~\\ \textbf{Notes} B2011\,=\,\citet{Bianchi2011}; S2019\,=\,\citet{TESS}; {Gaia}\,DR2\,=\,\citet{GaiaDR2}; W2010\,=\,\citet{WISE};\
D2015\,=\,\citet{Dunham2015}; R2017\,=\,\citet{Ribas2017}; P2016\,=\,\citet{Pascucci2016}.
\label{Tab:SED}
\end{table}

\begin{figure}
\begin{center}
\hspace{0cm}
\includegraphics[width=9cm]{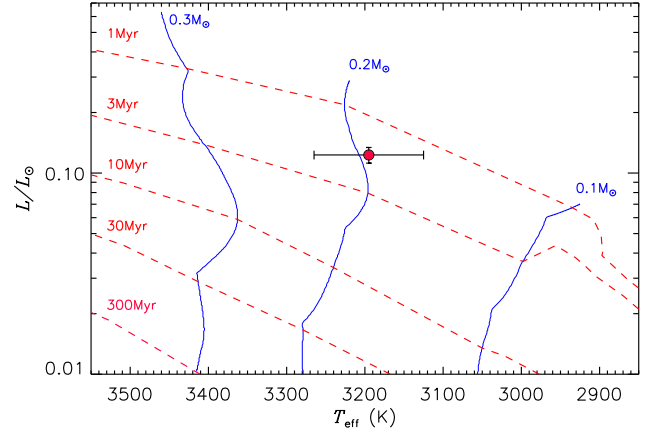}    
\vspace{0cm}
\caption{ISO-ChaI\, 52 in the Hertzsprung-Russell diagram. See Table\,\ref{Tab:param} for the values of \teff\ and $L_{\ast}$ derived
with the \ROTFIT\ code.  
Isochrones and evolutionary tracks by \citet{Baraffe2015} are overplotted  as dashed and solid lines,  with the labels representing their age
and mass, respectively. 
}
\label{fig:HR}
\end{center}
\end{figure}

\section{Additional plots}
\label{Appendix:plots}

\begin{figure}
\begin{center}
\hspace{0cm}
\vspace{-.3cm}
\includegraphics[width=9.cm]{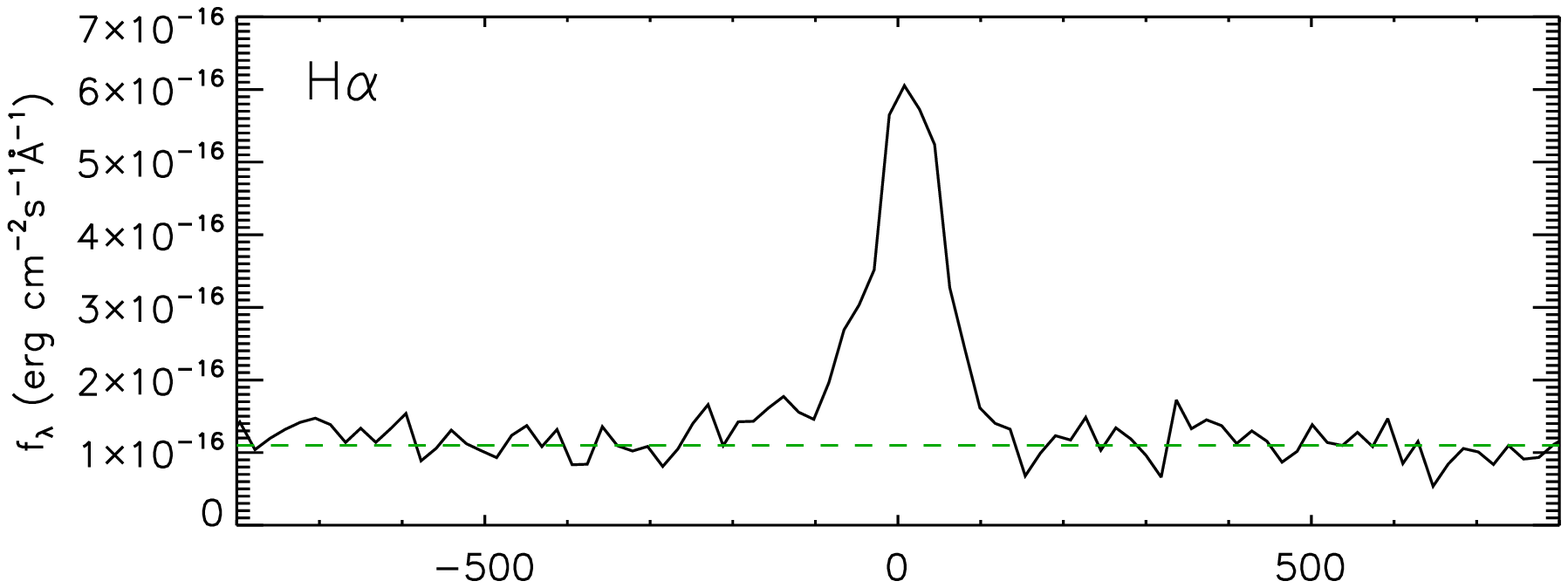} 
\vspace{-.3cm}
\includegraphics[width=9.cm]{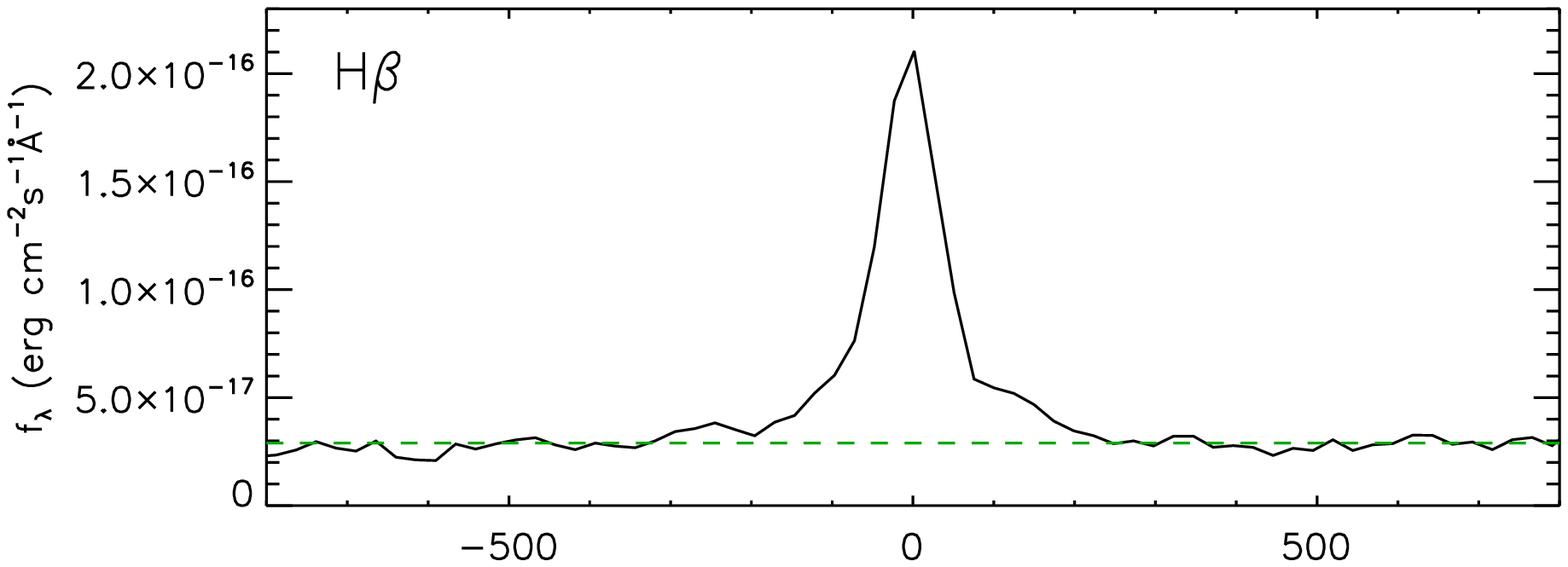} 
\vspace{-.3cm}
\includegraphics[width=9.cm]{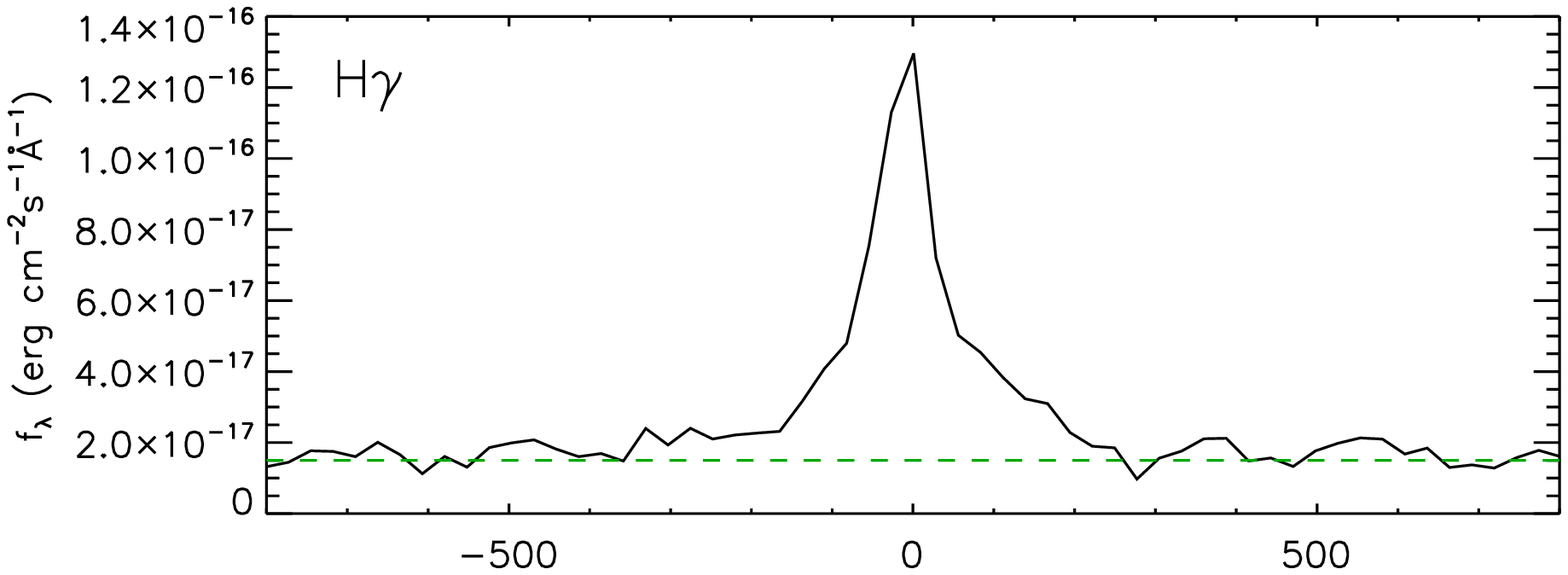} 
\vspace{-.3cm}
\includegraphics[width=9.cm]{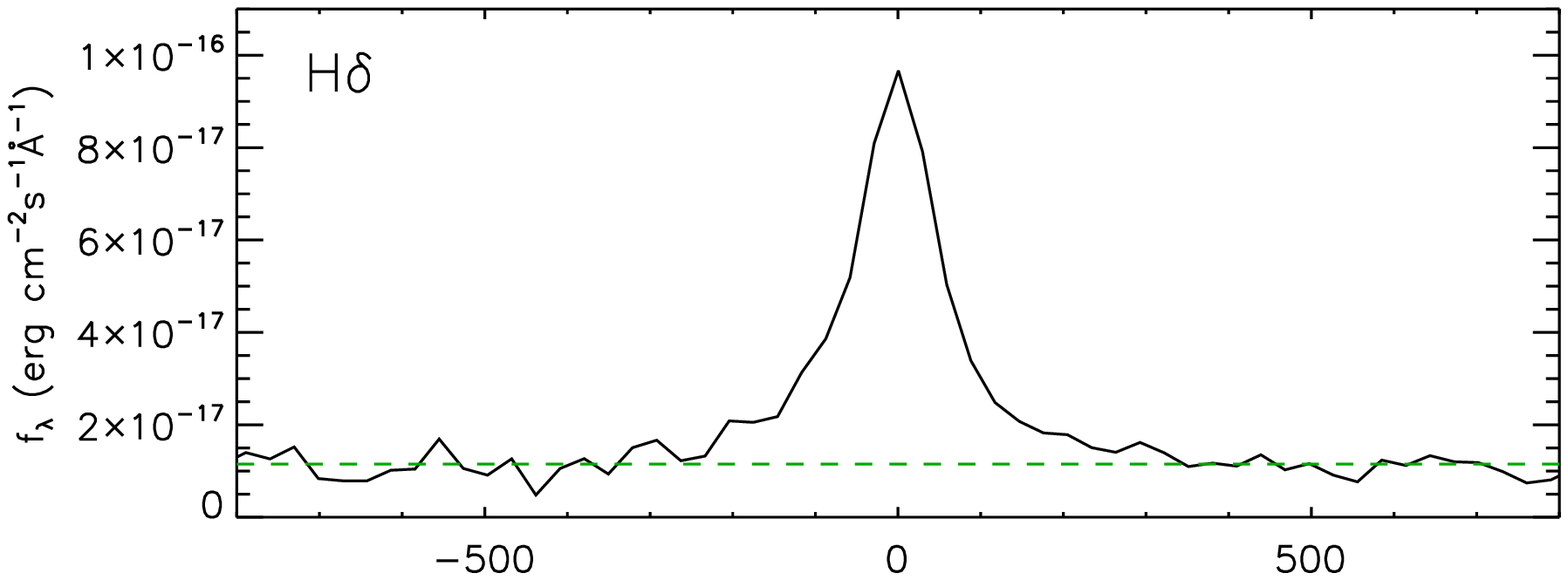}  
\vspace{-.3cm}
\includegraphics[width=9.cm]{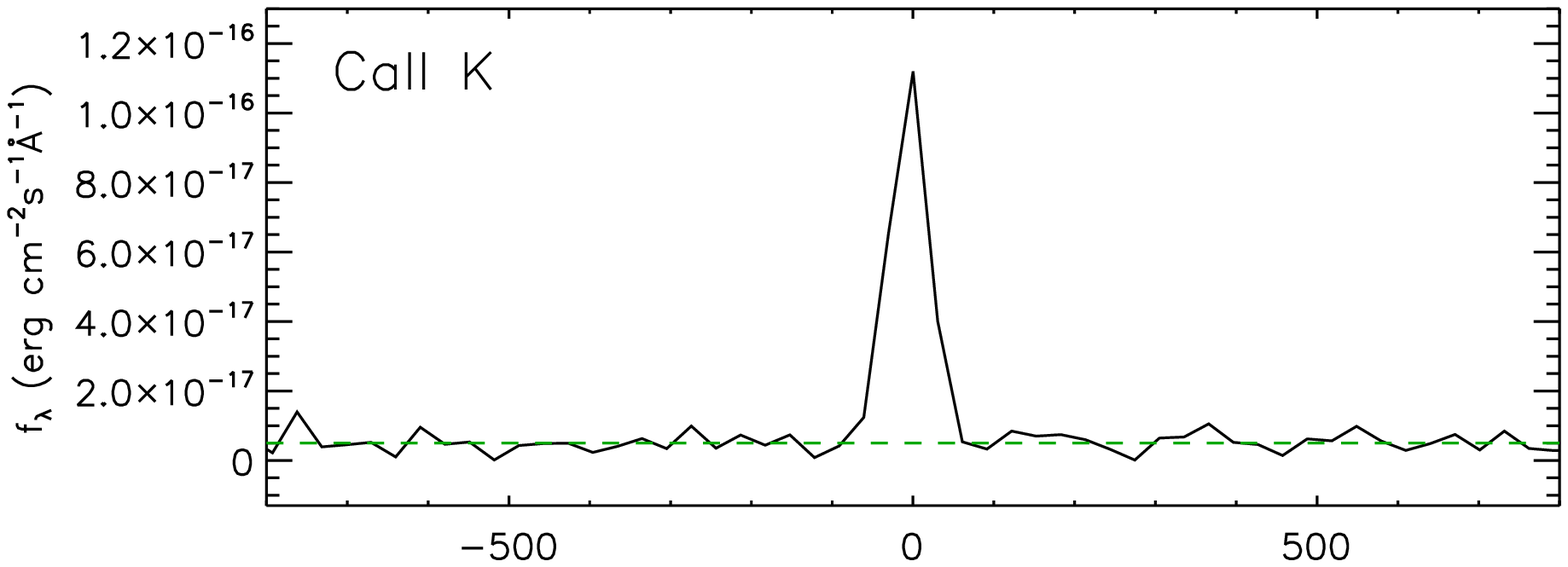} 
\vspace{-.3cm}
\includegraphics[width=9.cm]{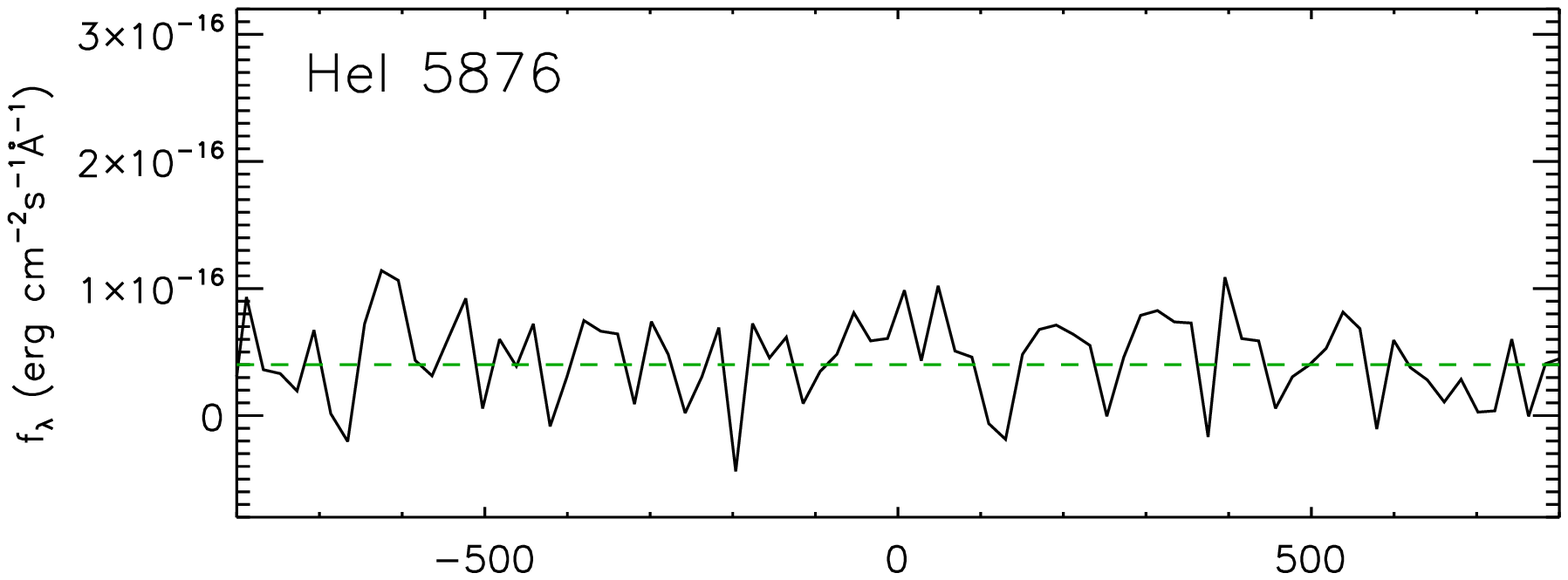} 
\vspace{0cm}
\includegraphics[width=9.cm]{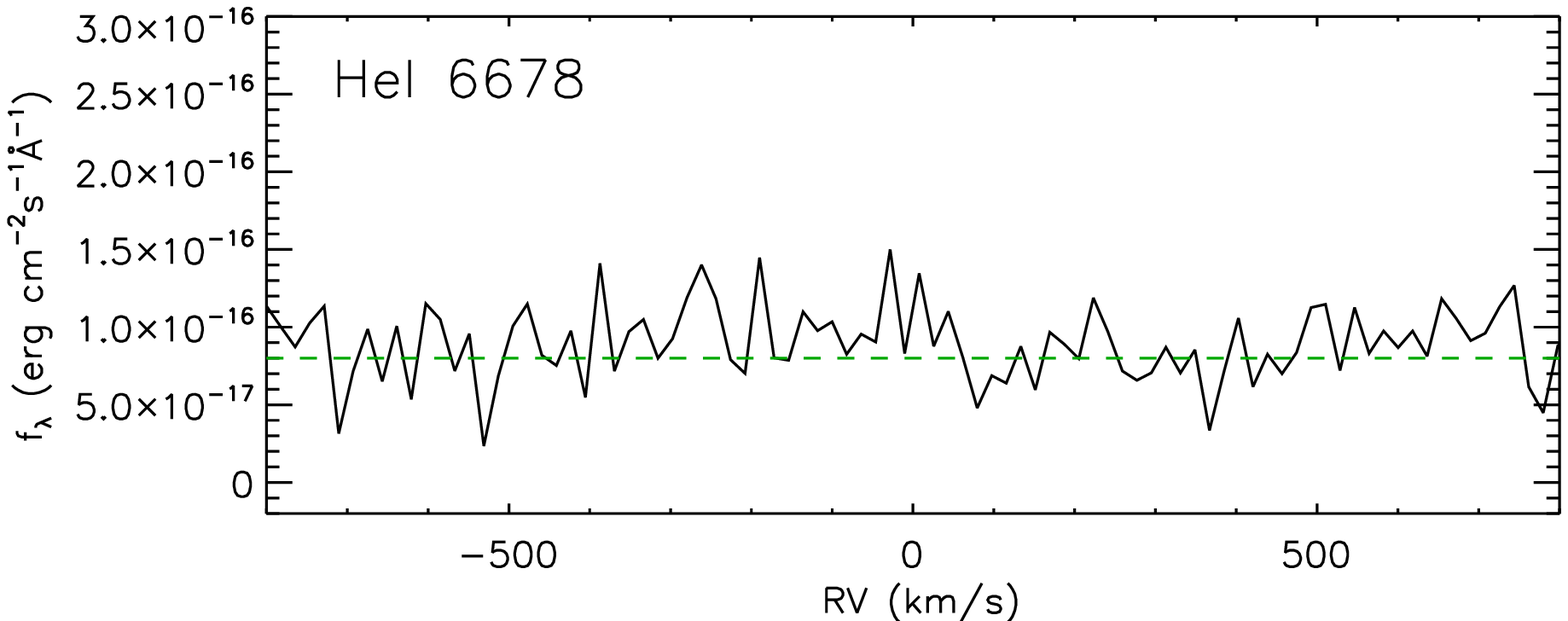}  
\caption{Profiles of Balmer lines, \ion{Ca}{ii}\,K, and \ion{He}{i} lines. In each plot, the horizontal dashed green line denotes the continuum level.}	
\label{fig:profiles}
\end{center}
\end{figure}

\begin{figure}
\hspace{-.5cm}
\includegraphics[width=9.5cm]{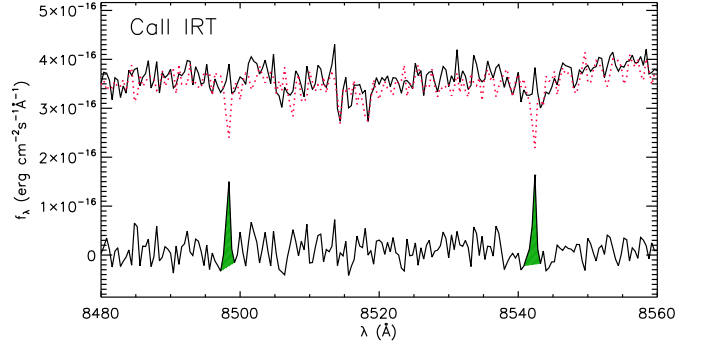}  
\vspace{0cm}
\caption{X-Shooter spectrum of ISO-ChaI\,52 in the region of \ion{Ca}{ii}\,IRT (solid black line) along with the inactive template
(dotted red line). The difference between observed  and  template  spectrum  is  shown  in  the  bottom of the box, along with the 
residual emission in the line cores (hatched green areas).
}
\label{fig:CaIRT}
\end{figure}

\begin{figure}
\begin{center}
\hspace{0cm}
\includegraphics[width=7.5cm]{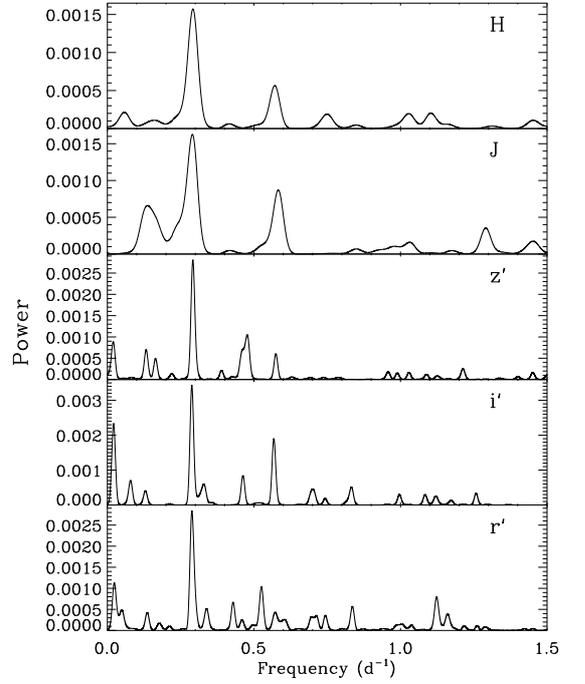} 
\vspace{0cm}
\caption{Cleaned periodograms for the photometric data of ISO-ChaI\,52 in the $HJz'i'r'$ bands ({\it from top to bottom}).
The highest peak in each band corresponds to a period of about 3.45 days.
}
\label{fig:periodograms}
\end{center}
\end{figure}

\begin{figure}
\begin{center}
\hspace{0cm}
\includegraphics[width=8.5cm]{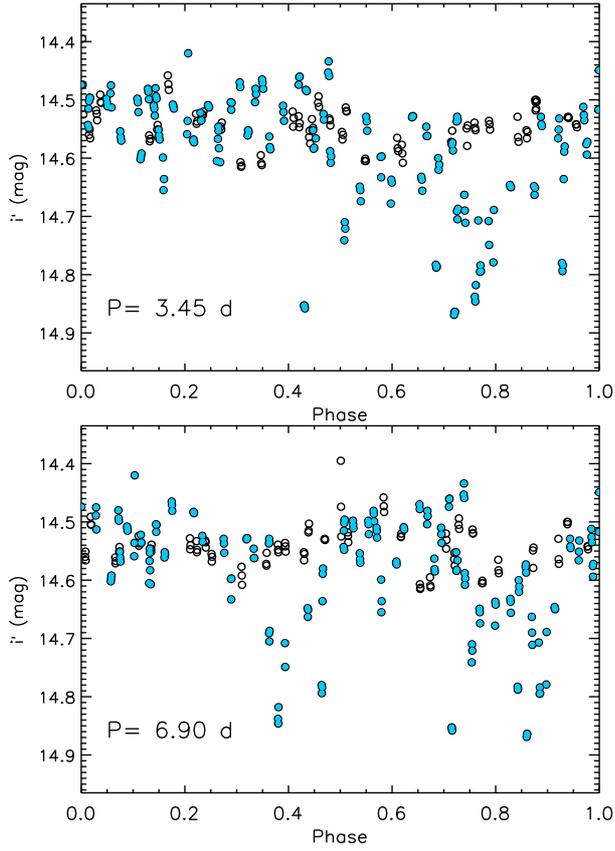} 
\vspace{0cm}
\caption{Light curve in the $i'$ band folded in phase with the period of 3.45 days ({\it upper panel}) and 6.9 days ({\it lower panel}). Filled dots refer to 
the data acquired before the end of May 2019 (JD$<2458630$); open dots are related to the second part of the data (Aug-Oct 2019).}
\label{fig:phased_lc}
\end{center}
\end{figure}

\end{document}